\documentclass[a4paper]{jpconf}
\usepackage{graphicx}
\usepackage{footmisc}
\usepackage{color}

\def\beq{\begin{equation}}
\def\eeq{\end{equation}}
\def\td#1{\tilde{\delta}\left(#1\right)}
\def\Eq#1{Eq.~(\ref{#1})}
\def\ep{\epsilon}
\def\qb{\mathbf{q}}
\def\beqn{\begin{eqnarray}} 
\def\eeqn{\end{eqnarray}}

\def\bea{\begin{eqnarray}}
\def\eea{\end{eqnarray}}
\def\nn{\nonumber}

\def\xiu{\xi_{1,0}}
\def\xid{\xi_{2,0}}

\def\v{{\rm V}}

\newcommand\as{\alpha_{\mathrm{S}}}

\def\uv{{\rm UV}}

\renewcommand{\thefootnote}{\fnsymbol{footnote}}

\begin{document}
\title{NLO cross sections in 4 dimensions without DREG}

\author{R. J. Hern\'andez-Pinto$^{1}$\footnote[1]{\hspace{1mm}Speaker}, F. Driencourt-Mangin$^2$, G. Rodrigo$^2$ and G. F. R. Sborlini$^2$}

\address{$^1$ Facultad de Ciencias F\'isico-Matem\'aticas, Universidad Aut\'onoma de Sinaloa, Ciudad Universitaria, CP 80000, Culiac\'an, Sinaloa, M\'exico.}

\address{$^2$ Instituto de F\'isica Corpuscular, Universitat de Val\`encia - Consejo Superior de Investigaciones Cient\'ificas, Parc Cient\'ific, E-46980 Paterna, Valencia, Spain.}

\ead{roger@uas.edu.mx}

\begin{abstract}
In this review, we present a new method for computing physical cross sections at NLO accuracy in QCD without using the standard Dimensional Regularisation. The algorithm is based on the Loop-Tree Duality theorem, which allow us to obtain loop integrals as a sum of phase-space integrals; in this way, transforming loop integrals into phase-space integrals, we propose a method to merge virtual and real contributions in order to find observables at NLO in $d=4$ space-time dimensions. In addition, the strategy described is used for computing the $\gamma^*\to q\bar{q}(g)$ process. A more detailed discussion related on this topic can be found in Ref~\cite{LTD0}.
\end{abstract}

\section{Introduction}
Theoretical tools for describing physical observables at the LHC require to be very precise for the determination of the properties of new particles and the description of new physics processes. In general, the formalism used to obtain accurate theoretical predictions is based  in the so-called Dimesional Regularisation~\cite{DREG} (DREG). In DREG, real and virtual integrals are implemented in $d=4-2\ep$ instead of $d=4$ space-time dimensions and singularities are manifested as poles in the $\ep$ parameter. However, cancellation of singularities on {\it infrared-safe observables} is guarantee by the KLN~\cite{KLN} theorem. These ideas have been studied in terms of the substraction techniques~\cite{substraction} and the developments along them have had an enormous impact on the theoretical predictions for the LHC. In this document, we will describe a new method based on the Loop-Tree Duality (LTD) theorem~\cite{LTD0,LTD1,LTD2,LTD3,LTD4,LTD5,LTD6,LTD7,LTD8,LTD9,LTD10,LTD11,LTD12}; we will present the general algorithm for the mapping between real and virtual variables so the cancelation of singularities is achieved at integrand level, meaning that integrals can be done in $d=4$ dimensions; finally, we will present a realisation of the method in the $\gamma^* \to q\bar{q}(g)$ process at NLO in QCD.

\section{The Loop-Tree duality}
LTD presents a new paradigm for writing loop integrals, at one-loop and beyond, as a sum of phase-space integrals, the so-called dual integrals. 
In order to illustrate the theorem at one-loop, let's consider a generic $N$-particle scalar one-loop integral with massive particles, i.e. 
\beq
L^{(1)}(p_1, \dots, p_N) = \int_{\ell} \, 
\prod_{i = 1}^N \,G_F(q_i)~,
\label{oneloop}
\eeq 
where $G_F(q_i) = (q_i^2-m_i^2+\imath 0)^{-1}$. Then, the dual representation consists of the sum of $N$ dual integrals:
\beq
L^{(1)}(p_1, \dots, p_N) 
= - \sum_{i = 1}^N \, \int_{\ell} \, \td{q_i} \,
\prod_{j =1, \, j\neq i}^N \,G_D(q_i;q_j)~, 
\label{oneloopduality}
\eeq 
where $G_D(q_i;q_j) = (q_j^2 -m_j^2 - \imath 0 \, \eta \cdot k_{ji})^{-1}$ are dual propagators, $i,j$ label the available internal lines and $\tilde{\delta}(q_i) \equiv 2 \pi  \imath \, \theta(q_{i,0}) \, \delta(q_i^2 -m_i^2)$. Masses and momenta of the internal lines are denoted $m_i$ and $q_{i,\mu} = (q_{i,0},\mathbf{q}_i)$, respectively, where $q_{i,0}$ is the energy and $\qb_{i}$ are  the spatial components. 
Internal lines can be written in terms of the loop momentum $\ell$ and the outgoing four-momenta of the external particles $p_i$, as 
$q_{i} = \ell + k_i  , k_{i} = p_{1} + \ldots + p_{i}, $ where momentum conservation holds, i.e. $k_{N} = 0$. Finally, \Eq{oneloopduality} shows explicitly a relation between a one-loop integral with $N$ external legs to $N$ phase-space integrals, where the change in sign in the $\imath 0$ prescription has to be considered carefully. This procedure can also be used for topologies with more than one-loop and it has been studied previously~\cite{LTD3}. 

\section{Unsubtraction in $d=4$}
The general procedure for describing {\it next-to-leading order} (NLO) cross sections involves the calculation of real and virtual corrections to the Born cross-section. Lets consider a process with $m$ particles in the final state; the NLO cross-section is written as
\beq
\sigma^{{\rm NLO}} = \int_m d\sigma^{\rm (1,R)}_{\rm V} +\int_{m+1}d \sigma^{(1)}_{\rm R}~,
\eeq
where $d \sigma^{(1)}_{\rm R}$ denotes the real correction and  $d\sigma^{\rm (1,R)}_{\rm V}$ contains all the information related to the renormalised virtual correction; indeed, using the LTD, virtual correction can be written as
\beq
 d\sigma^{\rm (1,R)}_{\rm V}  = \sum_{i=1}^{N} \int_{\ell} 2\,{\rm Re} \langle \mathcal{M}_N^{(0)}\vert \mathcal{M}_N^{\rm (1,R)} ( \tilde{\delta}(q_i)) \rangle\, \mathcal{O}_N(\{ p_i \})~,
\eeq
where $\mathcal{M}_N^{(0)}$ represents the {\it leading order} (LO) amplitude with $N$-legs, $\mathcal{M}_N^{\rm (1,R)}$ is the renormalised one-loop scattering amplitude, and $\mathcal{O}_N(\{ p_i \})$ is a measure function which is defined for a given observable. It is worth mentioning that  $\mathcal{M}_N^{\rm (1,R)}$ also contains the self-energy contributions of external legs, even if they becomes zero in the massless case. Besides, the renormalised amplitude includes counter-terms which are needed in order to subtract UV singularities locally, this aspect is widely discussed in Ref.~\cite{LTD0}. On the other hand, the real correction is given by
\bea
\int_{m+1}d \sigma^{(1)}_{\rm R} = \sum_{i=1}^N \int_{m+1} \vert \mathcal{M}^{(0)}_{N+1}(q_i,p_i)\vert^2 \mathcal{R}_i(q_i,p_i)\mathcal{O}_{N+1}(\{ p'_j \})~,
\label{eq:real}
\eea
where the external momenta $p_j'$, the phase-space and the tree-level scattering amplitude, $\mathcal{M}_{N+1}^{(0)}$, are rewritten in term of the loop three-momentum and the external momenta $p_i$ of the Born process.

One of the basic ideas in order to build a four dimensional representation of the scattering amplitudes resides on a proper mapping between real and virtual variables in order to merge both contributions at integrand level. Following these ideas, lets consider the first parton as the {\it emitter} and the second as the {\it spectator}; then, in order to reconstruct the kinematics associated to the real emission, we propose to map the loop three-momentum and the four momenta of the emitter and the spectator into a region of the real phase-space where the twin of the emitter decays to two partons in a soft and collinear configuration; it means,
\bea
\begin{array}{lllll}
 p_r'^\mu = q_i^\mu,\quad & 
 p_i'^\mu = p_i^\mu - q_i^\mu + \alpha_i \, p_j^\mu,\quad & 
 p_j'^\mu = (1-\alpha_i) \, p_j^\mu,\quad & 
 p_k'^\mu = p_k^\mu,\quad & k \ne i,j~,
\label{eq:multilegmapping}
\end{array}
\eea
where $\alpha_i = (q_i-p_i)^2/(2 p_j\cdot(q_i-p_i))$, $p_i$ is the momenta of the final-state emitter, $q_i$ is the internal on-shell momentum prior to the emitter, $p_j$ is the momentum of the final-state spectator, and $p_r'$ is the momentum of the extra radiated particle. In addition, it is straightforward to show that momentum conservation is satisfied in the mapping. A final piece in the method consists on the segmentation of the real phase-space. In \Eq{eq:real}, we have introduced a complete partition of the real phase-space
\bea
\sum \mathcal{R}_i (p_i,q_i) = \sum \prod_{jk\neq ir} \theta(y'_{jk} -y'_{ir}) = 1~,
\eea
which is similar to divide the phase-space as a function of the minimal dimensionless two-body invariants $y'_{ir} = s'_{ir}/s$. This separation of the real phase-space takes into account the relation between the real and virtual kinematics, such that, the real correction is actually contributing over the region where the soft and collinear divergences are generated. Hence, IR singularities are cancelled at the integrand level when both real and virtual corrections are simultaneously added. On the other hand, the UV divergences are subtracted locally by suitable counter-terms~\cite{LTD0}. Finally, the implementation of the NLO cross-section can be implemented in four-dimensions due to the absence of divergences at integrand level, i.e. $\ep=0$ in DREG.

\section{Unsubtraction applied to $\gamma^* \to q\bar{q} (g)$ at NLO in QCD}
In this section we present an example where the method can be tested. NLO correction to the process $\gamma^* \to q\bar{q}(g)$ includes, for the virtual correction, the self-energy and the triangle topologies, the most simple topologies where the LTD can be applied. Considering that internal loop momenta are $q_1=\ell + p_1$, $q_2=\ell + p_{12}$ and $q_3 = \ell$, we parametrise them as $2 q_i^{\mu} = \sqrt{s_{12}} \xi_{i,0} (1,2\sqrt{v_i(1-v_i)}{\bf{e}}_{i,\perp}, 1-2v_i) $ when $q_i^2 = 0$. Then, proceeding with the method described in the previous section, we obtain by combining real and virtual correction the dual cross-sections
\beqn
\nn \widetilde \sigma_{1}^{(1)} &=& \sigma^{(0)}  \frac{\as}{\pi}   C_F \, \int_0^1 d\xi_{1,0}  \int_0^{1/2} dv_1 \, {\cal R}_1(\xi_{1,0},v_1) 
\left[ 2 \left( \xiu - (1-v_1)^{-1} \right) - \frac{\xiu (1-\xiu)}{\left(1-(1-v_1)  \xiu \right)^2} \right],  
\label{eq:APPENDIXexplicitsigma1}
\\ \nn \widetilde \sigma_{2}^{(1)} &=& \sigma^{(0)}  \frac{\as}{2\pi}   C_F \, \int_0^1 d\xi_{2,0}  \int_0^1 dv_2  \, {\cal R}_2(\xi_{2,0},v_2) \, 
(1-v_2)^{-1}\,  \Bigg[ \frac{2 \, v_2 \, \xid \left(\xid (1-v_2) - 1\right)}{1-\xid}  \\ &&
- 1 + v_2\, \xid  + \frac{1}{1-v_2 \, \xid} \left(\frac{(1-\xid)^2}{(1-v_2 \, \xid)^2} + \xid^2\right)  \Bigg]~,
\label{eq:APPENDIXexplicitsigma2}
\eeqn
where the segmentation of the real phase-space is defined as
\bea
{\cal R}_1(\xiu,v_1) = \theta (1-2v_1) \, \theta\left( \frac{1-2v_1}{1-v_1} - \xiu \right), \quad 
{\cal R}_2(\xid,v_2) = \theta\left( \frac{1}{1+\sqrt{1-v_2}} - \xid \right)~;
\label{eq:errequeerre}
\eea
and the remnant dual pieces are collected in a single element leading to
\beqn
\nn \overline \sigma^{(1)}_{\v} &=& \sigma^{(0)} \, \frac{\as}{4\pi} \,  C_F \, \int_0^{\infty} d\xi \, \int_0^1 dv \, 
\Bigg\{ - 2 \, \left(1-{\cal R}_1(\xi,v)\right) \, v^{-1}(1-v)^{-1} \, \frac{\xi^2 (1-2 v)^2+1}{\sqrt{(1+\xi)^2 - 4 v\, \xi}} 
\\ \nn &+& 2 \, \left(1-{\cal R}_2(\xi,v)\right) \, (1-v)^{-1}\,  \left[2 \, v \, \xi \, \left(\xi (1-v) - 1\right)
\left(\frac{1}{1-\xi+\imath 0} + \imath \pi \delta(1-\xi)  \right) - 1 + v \, \xi \right] 
\\ \nn &+& 2\, v^{-1} \left( \frac{\xi (1-v)(\xi (1-2v)-1)}{1+\xi} +1\right) 
- \frac{(1-2v) \, \xi^3 \, ( 12 - 7 m_{\uv}^2 - 4 \xi^2 )} {(\xi^2 + m_{\uv}^2)^{5/2}} \\ 
&-& \frac{2\, \xi^2 (m_{\uv}^2 + 4\xi^2(1-6v(1-v)))}{(\xi^2 + m_{\uv}^2)^{5/2}} \Bigg\}~. 
\label{eq:APPENDIXexplicitremnant}
\eeqn 
In \Eq{eq:APPENDIXexplicitremnant} we have unified the integration variables, $\xi = \xi_{2,0} = \xi_{3,0} = \xi_{\rm UV}$ and $v=v_2=v_3=v_{\rm UV}$, while $(\xi_{1,0},v_1)$ are expressed in terms of $(\xi_{2,0},v_2)$ with the appropriate change of variables. The last two termns in \Eq{eq:APPENDIXexplicitremnant} are the UV counter-terms of the vertex and self-energies. Finally, integrating these expressions, we obtain the well known result
\bea
\sigma = \sigma^{(0)}\left(1+  3\,C_F \,\frac{\as}{4\pi} +\mathcal{O}(\as^2)\right)~,
\eea
where $\sigma^{(0)}= \alpha \, e_q^2 \, C_{\rm A}/2$. It is worth mentioning that using the method described in here, it was unnecessary to introduce any tensor reduction; Gram determinants are naturally avoided in LTD, and therefore the spurious singularities that the tensor reduction introduces leading to numerical instabilities in the integration over the phase-space.

\section{Conclusions}
In this review, we illustrated a new method for computing NLO cross sections with the application of the LTD theorem and without using DREG. The general algorithm is described and a practical realisation is implemented in the calculation of the $\gamma^*\to q\bar{q}(g)$ at NLO in QCD. The advantage on the use of the LTD is that the calculation of a full NLO correction is achieved with purely four-dimensional expressions, i.e. with the DREG parameter $\ep=0$.

\section{Acknowledgements}
The work of RJHP is partially supported by CONACyT, M\'exico. This work has been supported by CONICET Argentina, by the Spanish Government and ERDF funds from the European Commission (Grants No. FPA2014-53631-C2-1-P and SEV-2014-0398) and by Generalitat Valenciana under Grant No. PROMETEOII/2013/007. FDM acknowledges support from Generalitat Valenciana (GRISOLIA/2015/035).

\end{document}